\documentclass[useAMS,usenatbib]{mn2e}
\usepackage{graphicx}
\usepackage{multirow}
\usepackage{dcolumn}
\usepackage{array}

\title[]{Shape, Thermal and Surface Properties determination of a Candidate Spacecraft Target Asteroid (175706) 1996 FG3}
\author[LiangLiang Yu, Jianghui Ji \& Su Wang]{LiangLiang Yu$^{1,2}$, Jianghui Ji$^{1}$\thanks{jijh@pmo.ac.cn},
Su Wang$^{1}$\\
$^{1}$Key Laboratory of Planetary Sciences, Purple Mountain Observatory, Chinese Academy of Sciences, Nanjing 210008, China\\
$^{2}$Graduate School of Chinese Academy of Sciences, Beijing 100049, China}

\begin{document}
\date{Received 2013 November 14; in original form 2013 April 30}

\pagerange{\pageref{firstpage}--\pageref{lastpage}} \pubyear{2002}

\maketitle

\label{firstpage}

\begin{abstract}
In this paper, a 3D convex shape model of (175706) 1996 FG3, which
consists of 2040 triangle facets and 1022 vertices, is derived from
the known lightcurves. The best-fit orientation of the asteroid's spin axis is
determined to be $\lambda =237.7^\circ$ and $\beta=-83.8^{\circ}$ considering the
observation uncertainties, and its rotation period is $\sim$ 3.5935~h . Using the derived shape
model, we adopt the so-called advanced thermophysical model (ATPM)
to fit three published sets of mid-infrared observations of 1996 FG3
\citep{Wolters2011,Walsh2012}, so as to evaluate its surface
properties. Assuming the primary and the secondary bear identical
shape, albedo, thermal inertia and surface roughness, the best-fit
parameters are obtained from the observations. The geometric albedo
and effective diameter of the asteroid are reckoned to be $p_{\rm
v}=0.045\pm0.002$, $D_{\rm eff}=1.69^{+0.05}_{-0.02}$~km. The
diameters of the primary and secondary are determined to be
$D_{1}=1.63^{+0.04}_{-0.03}$~km and $D_{2}=0.45^{+0.04}_{-0.03}$~km,
respectively. The surface thermal inertia $\Gamma$ is derived to be
a low value of $80\pm40\rm~Jm^{-2}s^{-0.5}K^{-1}$ with a roughness
fraction $f_{\rm R}$ of  $0.8^{+0.2}_{-0.4}$. This indicates that
the primary possibly has a regolith layer on its surface, which is
likely to be covered by a mixture of dust, fragmentary rocky debris
and sand. The minimum regolith depth is estimated to be
$5\sim20\rm~mm$ from the simulations of subsurface temperature
distribution, indicating that 1996 FG3 could be a very suitable
target for a sample return mission.

\end{abstract}


\section{Introduction}
(175706) 1996 FG3 (hereafter, 1996 FG3) is a binary near-Earth
asteroid (NEA) belonging to the Apollo type, which has a very low
$\Delta$v value $\sim$ $5.16\rm~kms^{-1}$ \citep{Perozzi2001,Christou2003}.
This asteroid was originally chosen as the target for a proposed sample
return mission \citep{Barucci2012}, called MarcoPolo-R.

Eclipse events have been observed in the binary system 1996 FG3 in
optical wavelengths \citep{Pravec2000,Mot2000}. The period of mutual
orbit is $16.14\pm0.01\rm~hr$, and the diameter ratio
$D_{2}/D_{1}$ is $0.28^{+0.01}_{-0.02}$ \citep{Sch2009}. The primary
is assumed to be an oblate ellipsoid with a major axis ratio
$a/b\sim1.2$, while the secondary is prolate with a major axis
ratio about 1.4. Furthermore, \citet{Sch2009} showed the rotation
period of the primary is $\sim3.6\rm~hr$, and the mass density
$1.4^{+1.5}_{-0.6}$ g cm$^{-3}$, where 1996 FG3 is generally
classified as a complex C-type
\citep{Binzel2001,Binzel2012,Deleon2011,Deleon2013,Rivkin2013}. The optical
magnitude and phase coefficient are derived as $H=17.76\pm0.03$
and $G=-0.07\pm0.02$ by \citet{Pravec2006}, while \citet{Wolters2011}
used additional data to derive $H=17.833\pm0.024$ and $G=-0.041\pm0.005$.

\citet{Mueller2011} obtained an area-equivalent diameter of
$1.8^{+0.6}_{-0.5}\rm~km$ and a geometric albedo of
$p_{v}=0.04^{+0.04}_{-0.02}$ for 1996 FG3, based on thermal
observations (3.6 and $4.5\rm~\mu m$) from the "Warm Spitzer"
space telescope. Furthermore, \citet{Wolters2011} measured the
effective diameter and geometric albedo of the asteroid to be
$D_{\rm eff}=1.68\pm0.25\rm~km$, $p_{v}=0.046\pm0.014$, $\eta=1.15$
for a solar phase angle $11.7^{\circ}$ from an NEATM
\citep{Harris1998,Delbo2002} procedure.
In addition, using NEATM method, \citet{Walsh2012} also obtained
1996 FG3's effective diameter and geometric albedo of
$D_{\rm eff}=1.90\pm0.28\rm~km$, $p_{v}=0.039\pm0.012$, beaming
parameter $\eta=1.61\pm0.08$, for a phase angle $67.4^{\circ}$.

\citet{Walsh2008} modeled the formation of the binaries to be a
history of rotation acceleration due to the YORP mechanism.
Generally, the asymmetric reflection of sunlight and asymmetric
thermal emission from an asteroid's surface produces a net force and
a net torque. The net force causes the orbit of the asteroid to
drift, i.e., Yarkovsky effect, and the net torque alters its rotation
period and direction of its rotation axis, i.e., YORP effect
\citep{Rozitis2012}. The asymmetric shape of an asteroid,
as well as the existence of its finite rotation period and thermal
inertia, plays a major role in affecting Yarkovsky and YORP effects.
Thermal inertia is an important parameter that controls temperature
distribution over the surface and sub-surface of the asteroid, and
it is defined by $\Gamma=\sqrt{\rho c\kappa}$, where $\rho$ is the
mass density, $\kappa$ the thermal conductivity, and $c$ the specific
heat capacity. According to the definition, thermal inertia depends
mainly on the regolith particle size and depth, degree of compaction,
and exposure of solid rocks and boulders within the top few centimeters
of sub-surface \citep{Rozitis2011}. Therefore, thermal inertia
may act as a vital indicator to infer the presence or absence of loose
material on the asteroid's surface.

The thermal inertia of an asteroid may be evaluated by fitting mid-infrared
observations by applying a thermal model to reproduce mid-IR emission
curves. \citet{Lagerros1996} proposed a so-called thermophysical model (TPM),
which is adopted to calculate infrared emission fluxes as a function
of the asteroid's albedo, thermal inertia, correction factor and so on.
For example, we provide several values of $\Gamma$ for the targets of
complete/future asteroid missions. On the basis of TPM, \citet{Muller2005}
measured the thermal inertia of asteroid (25143) Itokawa to be
$\sim$ $750\rm~Jm^{-2}s^{-0.5}K^{-1}$. Again, \citet{Muller2011} showed
that the $\Gamma$ of 1999 JU3 is likely to be in the range of
$200\sim600\rm~Jm^{-2}s^{-0.5}K^{-1}$. The average thermal inertia of
1999 RQ36 is estimated to be $\sim$ $650\rm~Jm^{-2}s^{-0.5}K^{-1}$
\citep{Muller2012}. Recently, \citet{Rozitis2011} proposed an advanced
thermophysical model (ATPM) to extensively investigate the thermal nature
of the asteroid. Subsequently, \citet{Wolters2011} obtained a best-fit
thermal inertia $\Gamma=120\pm50\rm~Jm^{-2}s^{-0.5}K^{-1}$ for 1996 FG3
with ATPM.

The structure of the paper is as follows. Firstly, Section 2 gives
a brief introduction to the thermophysical model developed by
\citet{Rozitis2011}. In Section 3, we concentrate our study on deriving
a new 3D shape model for the primary of 1996 FG3 from the known
optical lightcurves \citep{Pravec2000,Mot2000,Wolters2011}. The
rotation period and spin axis of the primary are again updated in
this work but they appear to be slightly different from those of
\citet{Sch2009}. Subsequently, in Section 4, we independently
develop computation codes that duplicates ATPM \citep{Rozitis2011},
and then carry out extensive fittings using three sets of
mid-infrared observations of 1996 FG3 \citep{Wolters2011,Walsh2012}
to investigate the primary's surface physical properties, such as
the average thermal inertia, geometric albedo and roughness. The
results show that this asteroid may have a very rough surface, in
the meanwhile the thermal inertia and geometric albedo seem to be
relatively low, indicative of the existence of loose material or
regolith spreading over the asteroid's surface. Moreover, Section 5
deals with the global surface and sub-surface temperature
distribution at the aphelion and perihelion, and the minimum depth of
regolith layer over the very surface of the primary from the
simulations. In final, Section 6 presents the primary results of
this work and gives a concise discussion.

\section{Theory Method}
\subsection{Advanced Thermophysical Model}
An asteroid is assumed to be a polyhedron composed of N triangle
facets in ATPM. For each facet, the conservation of energy leads to
the surface boundary condition \citep{Rozitis2011}
\begin{eqnarray}
(1-A_{B})(s_{i}\psi_{i}F_{Sun}+F_{scat})+(1-A_{th})F_{red}=
\nonumber\\
\varepsilon\sigma T^{4}\big|_{z=0}+(-\kappa\frac{dT}{dZ})\Big|_{z=0} ,
\label{surbc}
\end{eqnarray}
where $A_{B}$ is the Bond albedo, $s_{i}$ indicates whether facet
$i$  is illuminated by the Sun, $\varepsilon$ is the thermal emissivity,
$\sigma$ the Stefan-Boltzmann constant, $\kappa$ the thermal
conductivity, and $z$ the depth below the surface, respectively.
$\psi_{i}$ means a function that returns the cosine of the sunlight
incident angle. $F_{Sun}$ is the integrated solar flux at the
distance of the asteroid, which can be approximated by
\begin{equation}
F_{Sun}=\frac{F_{\odot}}{r_{\odot AU}^{2}} ,
\end{equation}
where $F_{\odot}$ is the solar constant, about $1367.5Wm^{-2}$, and
$r_{\odot AU}$  is the heliocentric distance in AU. $F_{scat}$ and
$F_{red}$ are the total multiple-scattered and thermal-radiated
fluxes incident onto the facet from other facets, respectively.
$A_{th}$ is the albedo of the surface at thermal-infrared
wavelengths.

Temperature $T$ can be written as a function of time $t$ and depth
$z$, i.e., $T(t,z)$.  It can be described by  one-dimension (1D)
heat diffusion equation
\begin{equation}
\rho c\frac{\partial T}{\partial t}=\kappa\frac{\partial^{2}T}{\partial z^{2}} ,
\label{heat}
\end{equation}

For the asteroid, which may have a regolith layer over its surface,
the temperature below the surface regolith is supposed to be
constant. Thus an internal boundary condition could be given by
\begin{equation}
\frac{\partial T}{\partial z}\Big|_{z\rightarrow\infty}\rightarrow0 ~.
\label{inbc}
\end{equation}
Theoretically, the temperature distribution can be directly obtained
by solving equation (\ref{heat}) and  the boundary conditions of
equations  (\ref{surbc}) and (\ref{inbc}). However, the surface
boundary condition is a nonlinear equation, thus we will attempt to
solve them with numerical methods.

In order to simplify the solution of this problem, it is useful  to
introduce a standard transformation \citep{Spencer1989,Lagerros1996} as
follows:
\begin{eqnarray*}
x &= &\frac{z}{l_{s}} ,\\
\tau&=&\omega t ,\\
u&=&\frac{T}{T_{e}} ,
\end{eqnarray*}
where
\begin{displaymath}
l_{s}=\sqrt{\frac{\kappa}{\rho c\omega}}
\end{displaymath}
is referred to as skin depth. And

\begin{equation}
T_{e}=\left(\frac{(1-A_{B})F_{Sun}}{\varepsilon\sigma}\right)^{1/4}
~,
\end{equation}
is called effective temperature.

Thus, the 1D heat diffusion equation is rewritten as
\begin{equation}
\frac{\partial u}{\partial \tau}=\frac{\partial^{2}u}{\partial x^{2}} ,
\label{tf}
\end{equation}
while the surface boundary condition and internal boundary condition are converted into
\begin{equation}
u^{4}\big|_{x=0}-\Phi\frac{du}{dx}\Big|_{x=0}=p_{1}+p_{2}+p_{3} ,
\label{sbc}
\end{equation}
\begin{equation}
\frac{\partial u}{\partial x}\Big|_{x\rightarrow\infty}\rightarrow0
\label{ibc}
\end{equation}
where
\begin{eqnarray*}
p_{1}&=&s\cdot\psi ,\\
p_{2}&=&(1-A_{B})\frac{F_{scat}}{\varepsilon\sigma T^{4}_{e}} ,\\
p_{3}&=&(1-A_{th})\frac{F_{red}}{\varepsilon\sigma T^{4}_{e}} ,
\end{eqnarray*}
and
\begin{equation}
\Phi=\frac{\Gamma\sqrt{\omega}}{\varepsilon\sigma T^{3}_{e}}
\end{equation}
$\Gamma$ is the thermal inertia, and $\Phi$ is the thermal parameter.

By numerically solving equation (\ref{tf}) with boundary condition
equations (\ref{sbc}) and (\ref{ibc}), we can acquire the global
temperature distribution over the surface and sub-surface of the
asteroid. According to the theory, the program codes are
also developed for investigation. In the following, we will introduce
the results based on numerical calculations.

\subsection{Emission Flux}

\begin{figure*}
  \centering
\includegraphics[scale=0.5]{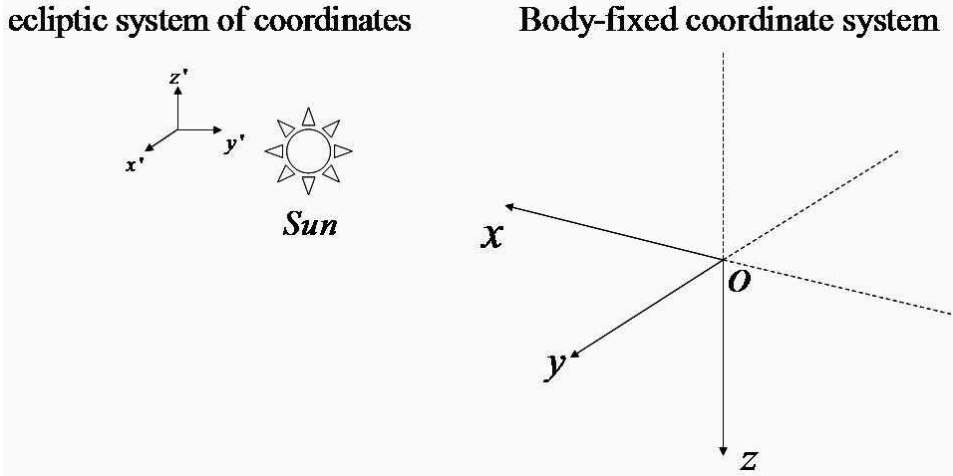}
  \caption{The coordinate system adopted in the model, where $O$ is the center of the asteroid,
  and z-axis is the spin axis.
  }\label{coordinate}
\end{figure*}

On the basis of surface temperature distribution calculated from ATPM, we may
estimate thermal infrared fluxes from the asteroid for a given phase angle
$\alpha$ and geocentric distance $\Delta$. First of all, a body-fixed
coordinate system is established (see Figure 1), where the origin $O$
locates at the center of asteroid, z-axis is parallel to its spin axis,
and x-axis is chosen to remain in a plane determined
by z-axis and the line of sun-asteroid, directing towards the Sun.

The direction vector of Earth in the body-fixed system is
calculated from the asteroid's orbit, expressed as
$\vec{n}_{\oplus}$. Normal vector of each facet $\vec{n}_{i}$ is
determined from the shape model of the asteroid. Hence, the view
factor of each facet relative to Earth can be evaluated by
\begin{equation}
f(i)=v(i)A(i)\frac{\vec{n}_{i}\cdot\vec{n}_{\oplus}}{\pi\Delta^{2}}
\end{equation}
where $v(i)=1$ indicates that facet $i$ can be seen from  Earth,
otherwise $v(i)=0$, and $A(i)$ is the area of facet $i$.

The flux emitted by each facet is described by the Plank function
\begin{equation}
B(\lambda, T_{i})=\frac{2\pi hc^{2}}{\lambda^{5}}\frac{1}{\exp\big(\frac{hc}{\lambda kT_{i}}-1\big)}
\end{equation}
Consequently, the total fluxes observed from Earth are fully integrated over each facet
\begin{equation}
F_{\lambda}=\sum^{N}_{i=1}\varepsilon f(i)B(\lambda, T_{i})
\end{equation}

\section{Shape Model}
According to direct images from space missions and radar measurements,
the asteroids in small-size have an irregular shape, while those
larger objects appear to be relatively regular shape. For instance,
Vesta seems to be more circular than Eros. Considering the
asteroid's rotation, the observed lightcurves change with a lot of
extrema, which provide key information on modeling its shape and
morphology. Hence, the substantial shape and spin status of the
asteroid can be derived from the observations. \citet{Kaa2001a} and
\citet{Kaa2001b}  developed the lightcurve inverse scheme, and in
their model, the inverse problem is perfectly solved using modern
deconvolution method and optimization techniques. With the
assistance of the optical data, a convex shape of asteroid will be
obtained. After repeating the fitting of the convex shape and
additional observations, an improved shape of the asteroid will be
better constructed.  The relative chi-square is defined as
\begin{equation}
\chi^2_{rel}=\sum_{i}\|\frac{L^{(i)}_{obs}}{\bar{L}^{(i)}_{obs}}-\frac{L^{(i)}}{\bar{L}^{(i)}}\|^2,
\end{equation}
where $L^{(i)}$ and $L^{(i)}_{obs}$ are the modeled and observed
brightness from lightcurves, respectively. $\bar{L}^{(i)}_{obs}$ and
$\bar{L}^{(i)}$  are the averaged brightness of the observed data
and the model, respectively. At the end of iteration, chi-square
will approach a tiny value. Usually, for an asteroid, there are a
great many of observations with respect to various solar angles at
different epochs. Therefore, such method will yield a reliable
result for the shape model.

\begin{table}\small
 \centering
  \caption{14 lightcurves adopted in the shape model of 1996 FG3.}
\label{lightcurve}
  \begin{tabular}{@{}cccl@{}}
  \hline
   ID & Obs. Date  & Obs. Number & References \\

 \hline
1 & 1998-12-13 & 155 & \citet{Mot2000} \\
2 & 1998-12-14 & 157 & \citet{Mot2000} \\
3 & 1998-12-15 & 156 & \citet{Mot2000} \\
4 & 1998-12-16 & 166 & \citet{Mot2000} \\
5 & 1998-12-10 & 111 & \citet{Pravec2000} \\
6 & 1998-12-17 & 110 & \citet{Pravec2000} \\
7 & 1999-01-09 & 237 & \citet{Pravec2000} \\
8 & 2009-03-25 & 16  & (MPC)\\
9 & 2009-03-27 & 14  & (MPC)\\
10& 2009-03-28 & 21  & (MPC)\\
11& 2009-03-29 & 20  & (MPC)\\
12& 2009-04-22 & 17  & (MPC)\\
13& 2011-01-29 & 35  & \citep{Wolters2011}\\
14& 2011-03-06 & 79  & \citep{Wolters2011}\\
 \hline
\end{tabular}
\end{table}

\begin{figure*}
  \centering
\includegraphics[scale=0.9]{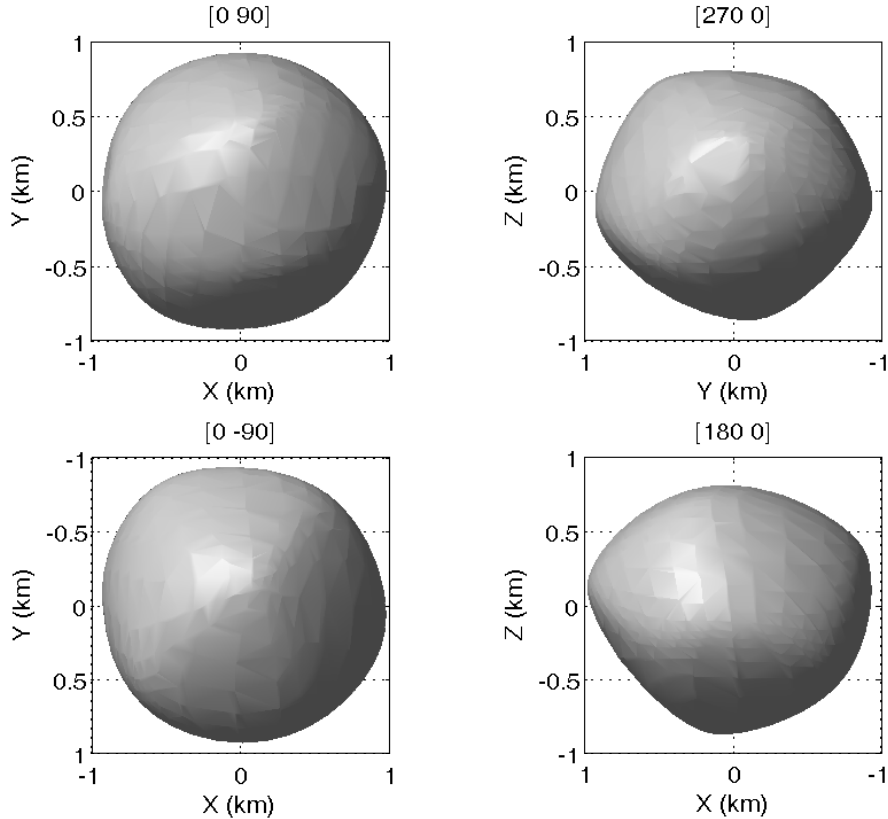}
  \caption{Derived shape model of 1996 FG3 shown from four view
  angles, i.e., north pole (top left), south pole (bottom
  left), equator (top right and bottom right), respectively.
  }\label{shape}
\end{figure*}

As 1996 FG3 is a binary asteroid, the values of the brightness in
the lightcurves consist of dual contributions coming from the primary
and secondary. Similar characteristic of the lightcurves is shown
by 1994 AW1 and 1991 VH \citep{Pravec1997,Pravec1998}. Thus, it is clear that
one cannot directly utilize the original data to construct the primary's shape model.
From the point of view of shape reconstruction, we should retrieve
the short-period component from the observations in the very beginning.
In the following, we describe the method that detaches the short-period
and long-period components.

By taking outside the deep minima from the long-period component,
the short-period component can be obtained. Then, applying the Fourier
fitting of the lightcurves to only short-period component, the
rotational profile for the primary can be detached
\citep{Mot2000,Pravec2000}. The Fourier series is in the form of
\citep{Harris1989,Pravec1996}

\begin{equation}
R(t)=C_0+\sum^m_{n=1}\left(C_n \cos\frac{2\pi n}{P}(t-t_0)+S_n
\sin\frac{2\pi n}{P} (t-t_0)\right),
\end{equation}
where $R$ is the magnitude computed at time $t$, the period
$P=3.5942\pm 0.0002$ h, $t_0$ is the epoch, $C_0$,  $C_n$ and $S_n$
are the mean magnitude and the Fourier coefficients of the $n$th
order, respectively. Subtracting the short-period component from the
observation data by assuming the long-period added to the
short-period linearly in irradiance, two kinds of components can be
gained.

\begin{figure*}
  \centering
\includegraphics[scale=0.6]{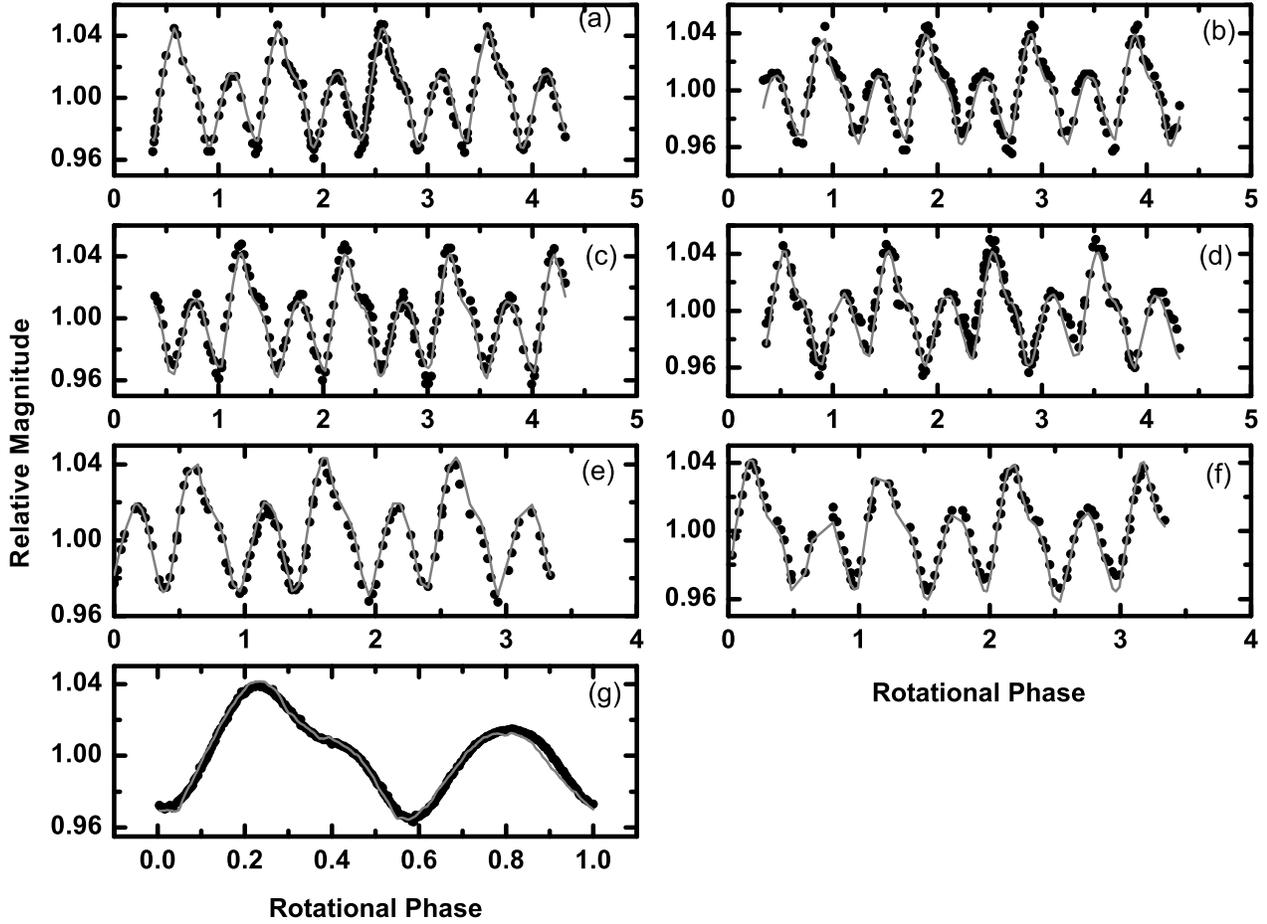}
  \caption{Comparison between the observed and modeling lightcurves.
  The red solid line means the modeling lightcurve while the black
  dot represents the observations. Panel (a) to (g) are the
  results of the cases with the first seven lightcurves (see Table \ref{lightcurve}).
  }\label{light}
\end{figure*}

\begin{figure*}
  \centering%
\includegraphics[scale=0.60]{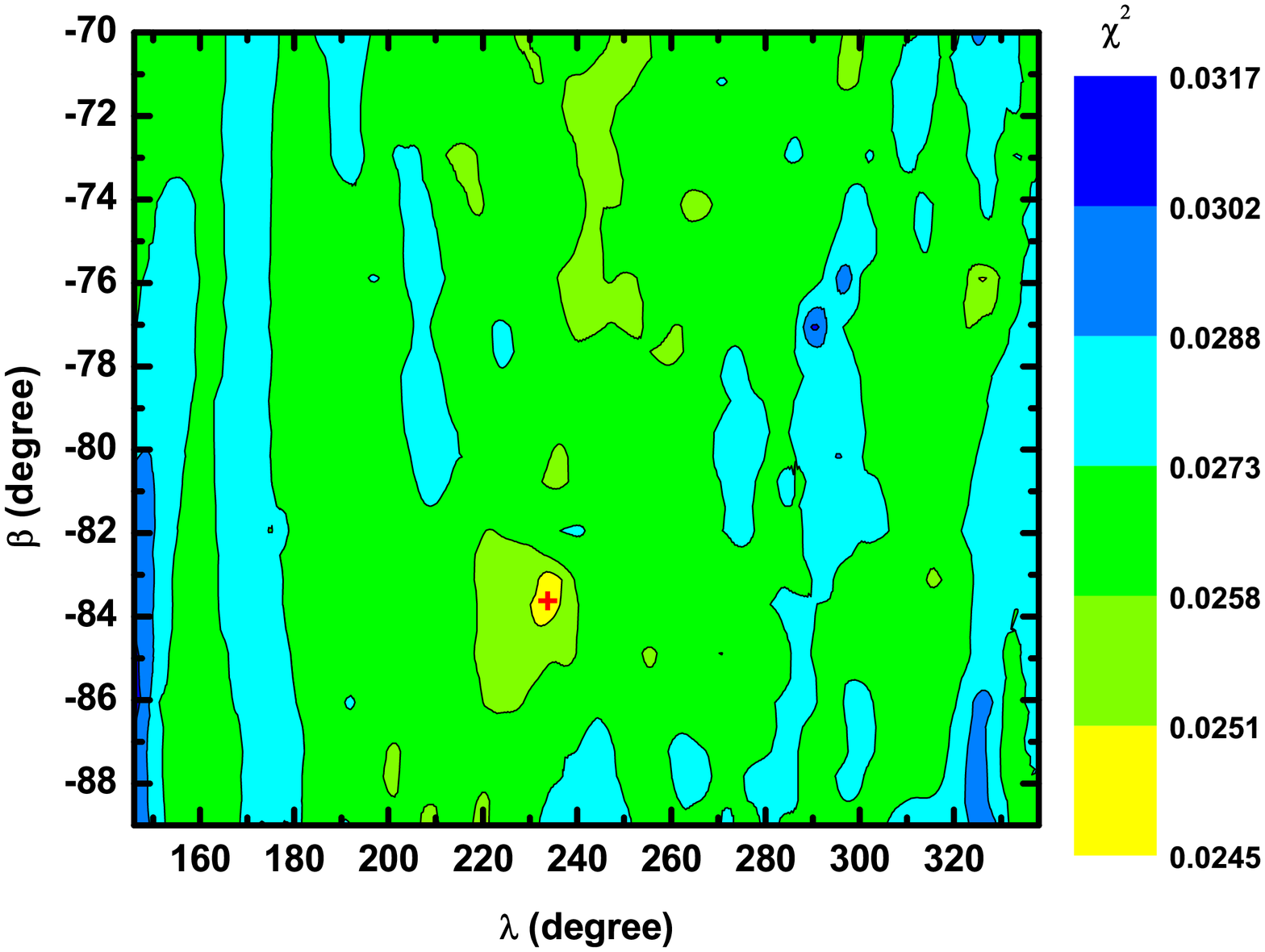}
  \caption{Contour of $\chi^2$ value. $\lambda$ and $\beta$ are
  chosen in their uncertainties of observed value. The red cross
  represents the pole orientation we determined from the shape model
  inversion process.
  }\label{chi}
\end{figure*}

By subtracting short-period data, the long-period measurements show
the information of eclipses/occultations which is not enough for the
shape reconstruction of the secondary. Herein, we simply make use of
the short-period data presented by the measurements that
have been published already to precisely reconstruct the shape of
the primary. We have already collected all optical data from 1996 to
now, where there are 14 lightcurves of 1996 FG3 (see Table
\ref{lightcurve}) in total
\citep{Mot2000,Pravec2000,Pravec2006,Sch2009,Wolters2011}.  From the
observation data of one to seven lightcurves, the best-fit
fifth-order Fourier series separates the short-period component and the
r.m.s residual is comparable to the observation errors. We can
obtain the short-period data of the first seven lightcurves, where
the first four lightcurves are adopted from Figure 1 of
\citet{Mot2000}, while the data of the fifth to seventh lightcurves
are from Figures 1 to 3 of \citet{Pravec2000}. The short-period
component of the first seven lightcurves has already obtained in the
published papers. In the observation interval of the remaining
lightcurves, no eclipsed or occultations occurred with an analysis
of amplitude of the data. Note that Table \ref{lightcurve}
summarizes all measurements that we employed in the shape model, the
data number in the table is simply referred to short-period data.
Among them, seven lightcurves (labled from No.1 to No.7) are in good
quality to perform the fitting to model 3D shape of this asteroid,
while the remaining data of each lightcurve from 8 to 14 are so
short in the observation time, less than one rotation period of the
primary, that we cannot get the short component from the total data,
especially No.8 to No.12. The short-period component of lightcurves
observed by \citet{Wolters2011} (No.13 and No.14) can be obtained
through Fourier fitting. However, we still use the data of No.13 and No.14
to perform additional investigation over the shape model, which is referred to
cases A1 and A2, respectively. The detailed analysis will be given in the end
of this section. In what follows we simply employ the data of the first seven
lightcurves for exploring the physical properties of 1996 FG3, including
the shape model inversion and the determination of the rotation period and
the pole orientation for the primary.

We deal with the lightcurve data as the input format for deriving a
shape model. All of data in the seven lighcurves are different, thus
we first convert them to relative brightness. Using the observation
epoch in JD, we calculate the ecliptic astrocentric cartesian
coordinates of the Sun and Earth. Firstly we determine the rotation
period of the primary by scanning in an interval of rotation period
space. Secondly, the rotation period is fixed at the determined
value and with free pole orientation parameters, the pole
orientation and shape model are obtained. At last, in order to
examine our results, we scan the $\chi^2$ value in the range of the
orientation uncertainties.

Based on optical observations, the rotation period of 1996 FG3
is estimated to be $\sim 3.595\pm 0.002$ h \citep{Mot2000}. Firstly,
we confirm this spin period of primary using the methods of
Kaasalainen et al. (2001 b) and \citet{Durech2010}. In addition, we
will find out a best-fit period on the basis of the published
lightcurves, i.e., in search of a global minimum of $\chi^2$ in the
fitting. In the parameter scanning process, six random pole
directions are chosen. Using Levenberg-Marquardt \citep{Press1992},
for each pole the shape model and rotation period are
simultaneously optimized to fit the observed lightcurves
\citep{Lowry2012}. In this work, we choose a scanning range of
period in [3.593, 3.597] h, based on the former outcomes
\citep{Mot2000,Pravec2000,Pravec2006}. Then, we carry out the
calculations after iterating 50 steps for each run, and have a
clearly minimum $\chi^2=0.005$ corresponding to the best-fitting
rotation period of 3.5935 h for the primary which is consistent
with that of \citet{Mot2000}.

Furthermore, to determine the asteroid's orientation, we then remain
the rotation period as 3.5935 h, but let the orientation be free
parameters. Herein, we adopt the initials of $\lambda =242^\circ$
and $\beta=-84^{\circ}$ \citep{Mot2000} to perform searching of
orientation. For the simulation, we choose the convexity
regularization weight to be 0.1, the laplace series expansion to be
$l=6$ and $m=6$, the light scattering parameters to be $a=0.5$,
$d=0.1$, $k=-0.5$, and $c=0.1$, where $a,~d,~k,~c$ means amplitude,
width, slope and Lambert's coefficient, respectively.

From this method, we are able to evaluate the orientation of 1996
FG3, the rotation period and a convex polyhedron shape model that
consists of the coordinates of the vertices and the number of the
facets of the polyhedron. After running many fittings, we finally
obtain one of the best-fit solutions for the shape model of 1996
FG3, which is composed of 1022 vertices and 2040 facets (Table \ref{phpa}). Figure
\ref{shape} shows the convex 3D shape from four view angles, north
pole (top left), south pole (bottom left), equator (top right and
bottom right), respectively. From Figure \ref{shape}, in the left
panels, we observe that north and south poles are both relatively
flat within current resolution. The right panels show the equatorial
views, where the equator of the asteroid seems to be most widest
region and the south pole looks to be narrower than the north pole.
And the left side is narrower than the right side shown in the
bottom right panel in Figure \ref{shape}. Thus from the shape model,
Table \ref{result} lists the initial conditions and the outcomes of
the simulation for its shape. According to the calculation, the
best-fit orientation of the target is determined to be $\lambda =237.7^\circ$
and $\beta=-83.8^{\circ}$ with the $\chi^2=0.025$. Figure
\ref{light} shows the comparison results of the observation data in
black dot and the modeled data in red solid line in panel (a) to
(g). In Figure \ref{light}, the x-axis means the rotation phase
and the y-axis represents the relative magnitude. From the figure, we
conclude that the first seven lightcurves fit is good.

\begin{table}
 \centering
  \caption{The initial conditions and outcomes of the parameters
  in our model. State means that whether the parameter is fixed or free in the simulation.}
  \label{result}
  \begin{tabular}{@{}cccccc@{}}
  \hline
    & orientation ($\lambda$) &  orientation ($\beta$) & rotation period \\
    & $(^{\circ})$ & $(^{\circ})$ & (hour) \\

 \hline
  initial & 242.0 & -84.0 & 3.5942 \\
  state & free & free & free\\
  result & 237.7 & -83.8 & 3.5935\\

 \hline
\end{tabular}
\end{table}

In order to ascertain  whether the orientation is the most likely
solution, we again scan the orientation in the range of its
uncertainty for $\lambda =242^{\circ} \pm 96^{\circ}$ and
$\beta={-84^{\circ}}^{+14^\circ}_{-5^\circ}$ \citep{Sch2009}. Figure
\ref{chi} shows the results, while the x-axis and y-axis represent
the orientation $\lambda$ and $\beta$. The contour shows the value
of $\chi^2$ by the colorbar index. In the figure, the yellow color
corresponds to a small value of $\chi^2$  while the blue color is
related to a larger value. The red cross in the figure represents
the very orientation that we obtain from the shape model inversion
programme. From the Figure, we can see that the value of
$\chi^2<0.0251$ locates at the position of $\lambda=[231^\circ,
238^\circ]$ and $\beta=[-85^\circ, -81^\circ]$, which is shown in
yellow region in Figure 4. The red cross just falls into the yellow
region. However, the chi-squared values are very low over the scanning pole
orientation space (Figure 4), indicating that the uncertainty in
the position of the pole appears to be high as shown by \citet{Sch2009}.
As a result, we may come to the conclusion that the derived orientation
is the best-fit solution in the range of the observation uncertainties.

Furthermore, we perform additional fitting and inverse the shape
models with all of fourteen lightcurves (case A1) in Table 1 and
eight lightcurves (case A2), which consist of the first seven and
No.14 lightcurves. In both cases, we apply the Fourier fitting to
deal with the observations given by \citet{Wolters2011}. We repeat
similar simulation process using fourteen lightcurves (case A1) and
eight lightcurves (case A2), respectively. In case A1, the rotation
period of the primary is determined to be 3.5943 h. Given the
unaltered period and free pole orientation, we retrieve the shape
model with the best-fit pole orientation at $\lambda=238.8^\circ$
and $\beta=-87.6^\circ$. In case A2, we employ eight lightcurves
that has been analyzed by Fourier fitting. In this case, the best-fit solution
of the spin period and pole orientation is determined to be 3.5947 h,
$\lambda=234.5^\circ$ and $\beta=-82.0^\circ$, respectively.
In comparison with three groups of solutions, we note that the outcomes
of pole orientation and spin period do not vary dramatically.
However, as the data in 2009 and 2011 appear to be sparse, simply covering
the primary's spin period much shorter than one rotational period,
thus it is hard to precisely derive the short-period component
using Fourier fitting with these data.
Therefore, in this work we finally utilize the first
seven lightcurves case to determine the shape model of 1996 FG3 (see
Figure 2).

Recently, Arecibo and Goldstone completed the radar measurements for
1996 FG3 \citep{Benner}, where both primary and the secondary are
revealed. Compared with the radar images, we find that our derived
shape model from optical observations bears a resemblance in some
directions \citep{Benner,Kaa2001a}, which gives an indication of
that the resultant shape of 1996 FG3 herein is reasonable. It is
worth noting that the equatorial ridge of the primary revealed by
the radar images is clearly seen in Figure \ref{shape}. Moreover,
the upper section in the bottom right panel seems to be consistent
with the radar images.

\section{Simulation}
\subsection{Observation Data}
In this work, three sets of thermal-IR data of 1996 FG3 are utilized,
summarized in Tables \ref{obs1}, \ref{obs2} and \ref{obs3}.
These three groups of published data are provided by \citet{Walsh2012} and
\citet{Wolters2011}, respectively. In addition, the observation
geometry at three epochs are also described in Table \ref{obsgeo},
which are adpoted to be input parameters in the calculation of
thermophysical model.

\begin{table}
 \centering
  \caption{MIRSI Observations of 1996 FG3 on 1 May 2009 \citep{Walsh2012}}
  \label{obs1}
  \begin{tabular}{@{}cccc@{}}
  \hline
  & UT &  wavelength & Flux\\
  &    & ($\mu m$)& ($10^{-14}\rm~Wm^{-2}\mu m^{-1}$) \\
 \hline
  & 05:51 & 11.6 & $2.75\pm0.07$\\
  & 05:53 & 11.6 & $2.87\pm0.07$\\
  & 05:54 & 11.6 & $3.01\pm0.07$\\
  & 05:59 & 8.7  & $3.40\pm0.08$\\
  & 06:01 & 8.7  & $3.06\pm0.08$\\
  & 06:10 & 18.4 & $1.59\pm0.14$\\
  & 06:17 & 9.8  & $3.00\pm0.13$\\
  & 06:20 & 9.8  & $3.27\pm0.15$\\
  & 06:23 & 9.8  & $3.01\pm0.14$\\
  & 06:25 & 11.6 & $2.75\pm0.06$\\
  & 06:27 & 11.6 & $2.69\pm0.07$\\
  & 06:30 & 8.7  & $2.98\pm0.08$\\
  & 06:33 & 8.7  & $2.74\pm0.09$\\
  & 06:36 & 8.7  & $3.02\pm0.08$\\
  & 06:43 & 18.4 & $1.59\pm0.14$\\
  & 06:48 & 11.6 & $2.44\pm0.07$\\
  & 06:51 & 9.8  & $2.80\pm0.15$\\
 \hline
\end{tabular}
\end{table}

\begin{table}
 \centering
  \caption{VISIR Observations of 1996 FG3 on 2 May 2009 \citep{Walsh2012}}
  \label{obs2}
  \begin{tabular}{@{}cccc@{}}
  \hline
  & UT &  wavelength & Flux\\
  &    & ($\mu m$)& ($10^{-14}\rm~Wm^{-2}\mu m^{-1}$) \\
 \hline
  & 00:28 & 11.88 & $2.53\pm0.05$\\
  & 00:56 & 8.59  & $2.95\pm0.05$\\
  & 01:05 & 18.72 & $1.12\pm0.06$\\
  & 01:14 & 8.59  & $2.77\pm0.05$\\ \hline
\end{tabular}
\end{table}

\begin{table}
 \centering
  \caption{VISIR observations of 1996 FG3 on 19 January 2011 \citep{Wolters2011}}
  \label{obs3}
  \begin{tabular}{@{}ccc@{}}
  \hline
  MJD-2455580 &  wavelength & Flux \\
  (days) & ($\mu m$)& ($10^{-15}\rm~Wm^{-2}\mu m^{-1}$) \\
 \hline
  0.14598 &  11.52 & $4.91\pm0.17$\\
  0.14838 &  11.52 & $4.63\pm0.16$\\
  0.15077 &  11.52 & $4.76\pm0.16$\\
  0.15473 &  8.70  & $4.96\pm0.28$\\
  0.15725 &  8.70  & $4.78\pm0.26$\\
  0.15966 &  8.70  & $5.41\pm0.28$\\
  0.16884 &  11.52 & $5.07\pm0.19$\\
  0.17124 &  11.52 & $4.89\pm0.17$\\
  0.17852 &  10.65 & $5.01\pm0.25$\\
  0.18108 &  10.65 & $5.92\pm0.26$\\
  0.18346 &  10.65 & $5.34\pm0.25$\\
  0.18930 &  11.52 & $5.07\pm0.17$\\
  0.19175 &  11.52 & $5.14\pm0.17$\\
  0.19467 &  12.47 & $4.48\pm0.24$\\
  0.19711 &  12.47 & $4.26\pm0.23$\\
  0.19949 &  12.47 & $4.70\pm0.24$\\
  0.20183 &  12.47 & $4.48\pm0.24$\\
  0.20421 &  12.47 & $4.68\pm0.24$\\
  0.20668 &  12.47 & $4.15\pm0.23$\\
  0.20962 &  11.52 & $4.43\pm0.16$\\
  0.21205 &  11.52 & $5.02\pm0.17$\\
  0.21575 &  8.70  & $5.55\pm0.30$\\
  0.21844 &  8.70  & $5.52\pm0.28$\\
  0.22179 &  8.70  & $5.30\pm0.27$\\
  0.22856 &  11.52 & $5.13\pm0.17$\\
  0.23097 &  11.52 & $4.91\pm0.16$\\
  0.23438 &  10.65 & $5.46\pm0.26$\\
  0.23701 &  10.65 & $5.43\pm0.25$\\
  0.23937 &  10.65 & $5.63\pm0.25$\\
  0.24282 &  11.52 & $5.13\pm0.17$\\
  0.24525 &  11.52 & $5.31\pm0.17$\\
  0.25203 &  12.47 & $4.74\pm0.23$\\
  0.25445 &  12.47 & $4.53\pm0.23$\\
  0.25683 &  12.47 & $4.41\pm0.23$\\
 \hline
\end{tabular}
\end{table}

\begin{table}
 \centering
 \caption{Observation geometry at three epochs.}
 \label{obsgeo}
 \begin{tabular}{@{}cccccc@{}}
 \hline
 Date & Heliocentric  & Geocentric  & Solar phase  \\
  time   & distance & distance & angle \\
 (UTC)&  (AU) & (AU) & $(^{\circ})$ \\
 \hline
 2009-5-01 & 1.057 & 0.1568 & 67.4 \\
 2009-5-02 & 1.053 & 0.1566 & 69.1 \\
 2011-1-19 & 1.377 & 0.4047 & 11.7 \\
 \hline
\end{tabular}
\end{table}

\subsection{Model Parameters}
For a thermophysical model like ATPM, several physical parameters
are required in the computation, such as the shape model, roughness,
albedo, thermal inertia, thermal conductivity, thermal emissivity,
heliocentric distance, geocentric distance, solar phase
angle and so on. As the orbit of 1996 FG3 is accurately measured
by optical observations, thus the heliocentric distance, geocentric
distance and phase angle are not difficult to obtain.

Although the shape model of 1996 FG3 is derived from its
lightcurves, we still do not know the actual physical size of 1996 FG3,
because the shape model simply shows the relative dimensions of the
asteroid. However, fortunately, the effective diameter $D_{\rm eff}$,
geometric albedo $p_{v}$, and absolute visual magnitude $H_{v}$ of
an asteroid can be evaluated by the following equation \citep{Fowler1992}:
\begin{equation}
D_{\rm eff}=\frac{1329\times 10^{-H_{v}/5}}{\sqrt{p_{v}}}~(\rm km) ~,
\label{Deff}
\end{equation}
thus if two of the parameters are available, the third is
easy to achieve.

However, the temperature distribution over the surface of an asteroid
depends mainly on rotation state, thermal inertia, albedo and roughness,
while the temperature distribution of sub-surface is greatly affected by
thermal conductivity. Therefore, we will make use of surface temperature
to derive a mean thermal inertia for 1999 FG3, and further to estimate a
profile for thermal conductivity, so as to obtain a more accurate subsurface
temperature distribution, then the regolith depth may be estimated
more accurately. All required parameters are summarized in Table \ref{phpa},
except free parameters. Herein, we actually have three free parameters ---
thermal inertia, albedo or effective diameter and surface roughness, which
are investigated in the fitting process.

\begin{table*}
 \centering
 \caption{Assumed physical parameters used in ATPM.}
 \label{phpa}
 \begin{tabular}{@{}ccc@{}}
 \hline
 Property & Value & References \\
 \hline
 Number of vertices           &      1022               & this work  \\
 Number of facets             &      2040               & this work  \\
 Primary shape (a:b:c)        &  1.276:1.239:1          & this work  \\
 Primary spin axis orientation &  $\lambda =237.7^\circ$
                                  $\beta=-83.8^{\circ}$ & this work  \\
 Primary spin period          &      3.5935 h           & this work  \\
 Secondary spin period        &      16.14 h            & \citep{Sch2009} \\
 $D_{2}/D_{1}$                & $0.28^{+0.01}_{-0.02}$  & \citep{Sch2009} \\
 Absolute visual magnitude    &     17.833              & \citep{Wolters2011} \\
 Slope parameter              &     -0.041              & \citep{Wolters2011} \\
 Emissivity                   &      0.9                & assumption \\
 \hline
\end{tabular}
\end{table*}

\subsection{Fitting Procedure}
\begin{table*}
 \centering
 \caption{ATPM fitting results to the observations.
 (The combined effective diameters $D_{\rm eff}$ are given in km)}
 \label{fitl2}
  \begin{tabular}{@{}|c|cc|cc|cc|cc|cc|cc|@{}}
 \hline
 Roughness &\multicolumn{12}{c|}{Thermal inertia $\Gamma$ ($\rm~Jm^{-2}s^{-0.5}K^{-1}$)} \\
 \cline{2-13}
 fraction & \multicolumn{2}{c|}{0} & \multicolumn{2}{c|}{50}
          & \multicolumn{2}{c|}{100} & \multicolumn{2}{c|}{150}
          & \multicolumn{2}{c|}{200} & \multicolumn{2}{c|}{300} \\
 $f_{\rm R}$ & $D_{\rm eff}$ & $L^{2}$ & $D_{\rm eff}$ & $L^{2}$
             & $D_{\rm eff}$ & $L^{2}$ & $D_{\rm eff}$ & $L^{2}$
             & $D_{\rm eff}$ & $L^{2}$ & $D_{\rm eff}$ & $L^{2}$ \\
 \hline
 0.00 & 1.647 & 340.8 & 1.674 & 1101.0 & 1.684 & 2089.5 & 1.694 & 2926.4 & 1.705 & 3571.4 & 1.729 & 4412.6 \\
 0.05 & 1.646 & 286.6 & 1.675 & 930.8 & 1.687 & 1900.0 & 1.699 & 2711.9 & 1.708 & 3401.4 & 1.732 & 4265.2 \\
 0.10 & 1.644 & 267.6 & 1.677 & 777.5 & 1.689 & 1719.5 & 1.704 & 2505.0 & 1.710 & 3235.0 & 1.734 & 4119.9 \\
 0.15 & 1.642 & 283.3 & 1.678 & 641.5 & 1.692 & 1548.3 & 1.709 & 2306.1 & 1.713 & 3072.3 & 1.736 & 3976.7 \\
 0.20 & 1.639 & 332.9 & 1.679 & 522.7 & 1.694 & 1386.5 & 1.714 & 2115.5 & 1.716 & 2913.5 & 1.738 & 3835.8 \\
 0.25 & 1.635 & 415.7 & 1.680 & 421.3 & 1.696 & 1234.2 & 1.719 & 1933.5 & 1.718 & 2758.5 & 1.741 & 3697.1 \\
 0.30 & 1.631 & 530.7 & 1.680 & 337.4 & 1.698 & 1091.6 & 1.723 & 1760.4 & 1.721 & 2607.5 & 1.743 & 3560.7 \\
 0.35 & 1.626 & 676.9 & 1.680 & 270.9 & 1.700 & 958.8 & 1.728 & 1596.4 & 1.723 & 2460.6 & 1.745 & 3426.6 \\
 0.40 & 1.621 & 852.9 & 1.680 & 221.7 & 1.702 & 835.9 & 1.732 & 1442.0 & 1.726 & 2317.7 & 1.747 & 3294.9 \\
 0.45 & 1.616 & 1057.7 & 1.680 & 190.0 & 1.703 & 723.0 & 1.736 & 1297.3 & 1.728 & 2179.0 & 1.749 & 3165.6 \\
 0.50 & 1.610 & 1289.9 & 1.679 & 175.4 & 1.704 & 620.1 & 1.740 & 1162.6 & 1.730 & 2044.5 & 1.751 & 3038.7 \\
 0.55 & 1.604 & 1548.1 & 1.678 & 177.9 & 1.706 & 527.4 & 1.744 & 1038.2 & 1.732 & 1914.3 & 1.753 & 2914.2 \\
 0.60 & 1.597 & 1830.9 & 1.677 & 197.4 & 1.707 & 444.9 & 1.747 & 924.3 & 1.735 & 1788.4 & 1.755 & 2792.2 \\
 0.65 & 1.590 & 2136.8 & 1.675 & 233.5 & 1.707 & 372.5 & 1.751 & 821.3 & 1.737 & 1666.9 & 1.757 & 2672.7 \\
 0.70 & 1.583 & 2464.5 & 1.674 & 286.1 & 1.708 & 310.5 & 1.754 & 729.2 & 1.739 & 1549.8 & 1.760 & 2555.7 \\
 0.75 & 1.575 & 2812.3 & 1.672 & 354.9 & 1.709 & 258.7 & 1.757 & 648.4 & 1.741 & 1437.3 & 1.761 & 2441.3 \\
 0.80 & 1.567 & 3178.8 & 1.670 & 439.5 & 1.709 & 217.2 & 1.760 & 579.0 & 1.743 & 1329.2 & 1.763 & 2329.5 \\
 0.85 & 1.559 & 3562.6 & 1.667 & 539.7 & 1.710 & 185.9 & 1.763 & 521.2 & 1.744 & 1225.7 & 1.765 & 2220.3 \\
 0.90 & 1.551 & 3962.2 & 1.665 & 655.1 & 1.710 & 164.9 & 1.765 & 475.3 & 1.746 & 1126.8 & 1.767 & 2113.8 \\
 0.95 & 1.542 & 4376.0 & 1.662 & 785.3 & 1.710 & 154.0 & 1.768 & 441.2 & 1.748 & 1032.6 & 1.769 & 2009.9 \\
 1.00 & 1.534 & 4802.9 & 1.659 & 929.8 & 1.710 & 153.3 & 1.770 & 419.3 & 1.750 & 943.1 & 1.771 & 1908.7 \\
 \hline
\end{tabular}
\end{table*}

As 1996 FG3 is a binary system, and the rotation period of the
secondary is 16.14~h, different from that of the primary, the flux of
the secondary is modeled independently in the fitting procedure.
The overall thermal flux predictions is a summation of that
from both the primary and the secondary. Actually, the consideration
of the flux of the secondary has no significant affect on the result,
despite a very slight influence on the effective diameter. Thus,
to simplify fitting process, we assume that the secondary
shares an identical shape model with the primary.

On the other hand, the observations do not spatially resolve 1996
FG3. In this sense, the ATPM-derived diameter $D_{\rm eff}$ is simply
considered to be an effective diameter of a sphere with the combined
cross-sectional area of two components. Thus, we suppose that
the component diameters $D_{1}$ and $D_{2}$ are related to $D_{\rm eff}$
via $D_{1}^{2}+D_{2}^{2}=D_{\rm eff}^{2}$ \citep{Walsh2012}.

Surface roughness could be modeled by a fractional coverage of
hemispherical craters, symbolized by $f_{\rm R}$, while the
remaining fraction, $1-f_{\rm R}$, represents a smooth flat surface
on the asteroid. In this work, we adopt a low resolution
hemispherical crater model that consists of 132 facets and 73
vertexes, following a similar treatment as shown in
\citet{Rozitis2011}. As well-known, the sunlight is fairly easier to
be scattered on a rough surface than a smooth flat region, thus the
roughness can decrease the effective Bond albedo. For the
above-mentioned surface roughness model, the effective Bond albedo
$A_{\rm eff}$ of a rough surface can be related to the Bond albedo
$A_{\rm B}$ of a smooth flat surface and the roughness $f_{\rm R}$
by \citep{Wolters2011}
\begin{equation}
A_{\rm eff}=f_{\rm R}\frac{A_{\rm B}}{2-A_{\rm B}}+(1-f_{\rm R})A_{\rm B}~.
\label{aeffab}
\end{equation}
On the other hand, the effective Bond albedo $A_{\rm eff}$ is related to
geometric albedo $p_{v}$ by
\begin{equation}
A_{\rm eff}=p_{v}q_{\rm ph}~,
\label{aeffpv}
\end{equation}
where $q_{\rm ph}$ is a phase integral that can be approximated by
\citep{Bowell}
\begin{equation}
q_{\rm ph}=0.290+0.684G~,
\label{qph}
\end{equation}
where $G$ is the slope parameter in the $H, G$ magnitude system of
\citet{Bowell},  we chose $H=17.833\pm0.024$, $G=-0.041\pm0.005$
\citep{Wolters2011} in our fitting process. Thus each surface roughness
and effective diameter leads to a unique $A_{\rm eff}$ and $A_{\rm B}$.

In order to simplify our modeling process, a set of thermal inertia
are selected in the range $0\sim500\rm~Jm^{-2}s^{-0.5}K^{-1}$. And
for each thermal inertia case, a series of surface roughness and
effective diameter are calculated to examine which may act as
best-fit parameters with respect to the observations.

As ATPM requires a Bond albedo $A_{B}$ for each facet element to
simulate temperature distribution, thus an $A_{B}=0.01$ is herein
assumed to be initials for two components of 1996 FG3. For each
thermal inertia $\Gamma$, effective diameter $D_{\rm eff}$ and
surface roughness $f_{\rm R}$ case, a flux correction factor $FCF$
is defined as \citep{Wolters2011}
\begin{equation}
FCF=\frac{1-A_{B,now}}{1-A_{B,initial}}~,
\end{equation}
where $A_{B,now}$ is calculated by inversion of equation
(\ref{aeffab}), to fit the observations, and then an error-weighted
least-squares fit is defined as \citep{Harris1998, Wolters2011}
\begin{equation}
L^{2}=\sum^{n}_{i=1}\Big(\frac{FCF(f_{\rm R},D_{\rm eff})F_{\rm model}
    -F_{\rm obs}(\lambda_{n})}{\sigma_{\lambda_{n}}}\Big)^{2}~,
\label{l2}
\end{equation}
which can be obtained to evaluate the fitting degree of our model results
to the observations. It should be noticed that the predicted model
flux $F_{\rm model}=F_{\rm model}(\Gamma,f_{\rm R},D_{\rm
eff},\lambda_{n})$ is a rotationally averaged profile, because the
rotation phase of 1996 FG3 was unknown at the time of
observations.

The fitting outcomes are summarized in Table \ref{fitl2}. In the
table, the $L^{2}$ values are relevant to each thermal inertia,
roughness fraction and effective diameter. Roughly speaking, the
$L^{2}$ values imply a thermal inertia in the range
$0\sim150\rm~Jm^{-2}s^{-0.5}K^{-1}$.

\begin{figure}
\includegraphics[scale=0.6]{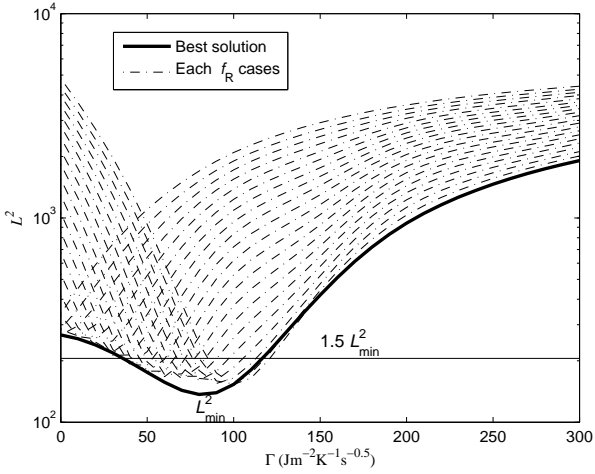}
  \centering
  \caption{$\Gamma\sim L^{2}$ profile fit to the observations.
  Each dashed curve represents a roughness fraction $f_{\rm R}$ in the
  range of $0.0\sim1.0$. The heavy line is a cubic spline interpolation
  curve for each lowest $L^{2}$ derived from each free parameter.
  }\label{frl2}
\end{figure}

To acquire a likely solution from Table \ref{fitl2}, the minimum
error-weighted least-squares $L^{2}$ value need to be determined and
an uncertain range of the minimum $L^{2}$ is then taken into account.
Firstly, the $\Gamma\sim L^{2}$ curves are drawn to understand how
$L^{2}$ alters with free parameters, including thermal inertia,
roughness fraction and effective diameter according to Table \ref{fitl2}
(see Figure \ref{frl2}). Secondly, the minimum $L^{2}$ is determined from
a cubic spline interpolation curve for each lowest $L^{2}$ and the related
free parameter in Figure \ref{frl2}. Furthermore, the minimum
$L^{2}$, symbolized as $L_{\rm min}^{2}$, arises at the case $f_{\rm R}=0.8$,
$\Gamma=80\rm~Jm^{-2}s^{-0.5}K^{-1}$ and $D_{\rm eff}=1.69\rm~km$.
As the outcomes are derived from a combined fit to three observation
epochs, the $L^{2}$ profiles thereby vary in a relatively broader range.
If a range of $50\% ~ L_{\rm min}^{2}$  is assumed, then
a significant uncertainty can be obtained.
Hence, the final adopted best-fit parameters for 1996 FG3 are
summarized in Table \ref{outcome}.

\begin{table}
 \centering
 \caption{Derived properties of 1996 FG3 from ATPM.}
 \label{outcome}
 \begin{tabular}{@{}cccc@{}}
 \hline
 Property                                               & Result\\
 \hline
 Thermal inertia $\Gamma$($\rm~Jm^{-2}s^{-0.5}K^{-1}$)  & $80\pm40$ \\
 Roughness fraction $f_{\rm R}$                         & $0.8^{+0.2}_{-0.4}$\\
 Geometric albedo $p_{v}$                               & $0.045\pm0.002$  \\
 Effective diameter $D_{\rm eff}$(~km)                  & $1.69^{+0.05}_{-0.02}$\\
 Primary diameter $D_{1}$(~km)                          & $1.63^{+0.04}_{-0.03}$\\
 Secondary diameter $D_{2}$(~km)                        & $0.45^{+0.04}_{-0.03}$\\
 \hline
\end{tabular}
\end{table}

In the following section, we will then utilize the above derived
parameters to evaluate the surface thermal environment of 1996 FG3
at its aphelion and perihelion respectively.

\section{Surface Thermal Environment}
\subsection{Temperature Distribution}
\begin{figure*}
\includegraphics[scale=0.66]{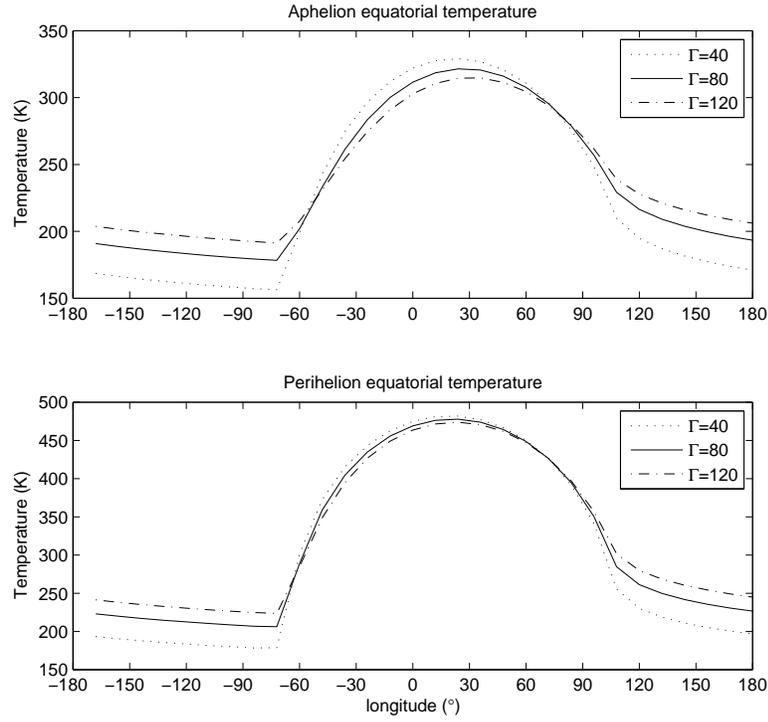}
  \centering
  \caption{Equatorial temperature distribution of 1996 FG3
  at aphelion and perihelion, respectively. Maximum temperature at $24^{\circ}$,
  minimum temperature at $-72^{\circ}$. ($0^{\circ}$ is the sub-solar point.)
  }\label{aph-per}
\end{figure*}

\begin{figure*}
\includegraphics[scale=0.6]{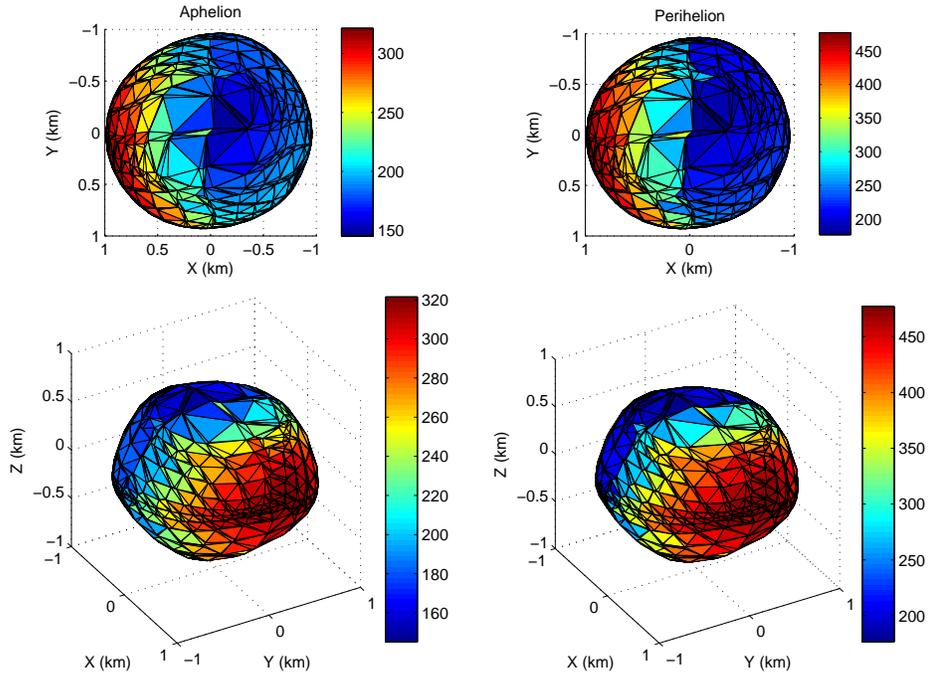}
  \centering
  \caption{Global surface temperature distribution obtained from
  ATPM at the aphelion and perihelion, respectively. The color bar
  indicates the range of temperature, where red
  for high temperature and blue for low temperature.
  }\label{gtemp}
\end{figure*}

On the basis of the derived shape model from the observations, the
physical parameters in Table \ref{phpa}, and the thermal inertia
determined from the ATPM fitting process, we attempt to simulate the
temperature variation of 1996 FG3 over a rotation period. In Figures
\ref{aph-per} and  \ref{gtemp}, we show the equatorial temperature
and global surface temperature distribution for the primary at
aphelion and perihelion, respectively.

Figure \ref{aph-per} shows the equatorial temperature distribution
of 1996 FG3 at its aphelion and perihelion respectively. The maximum
temperature does not appear at the solar point, but delays about
$24^\circ$, and the minimum temperature occurs just a little after the
local sunrise, delaying about $18^\circ$. Such delay effect between
absorption and emission is actually caused by the non-zero thermal
inertia and the finite rotation speed of the asteroid.
On the other hand, according to Figure \ref{aph-per}, the
equatorial temperature of 1996 FG3 may range from
$180$ to $480\rm~K$ over a whole orbit period.

Figure \ref{gtemp} shows the global surface temperature distribution
of 1996 FG3  at its aphelion (left) and perihelion (right) respectively.
In this figure, z-axis represents the asteroid's spin axis, and x-axis
points to the Sun in the framework of an asteroid center body-fixed coordinate
system. The profile of temperature in Figure \ref{gtemp} is shown by
the index of color-bar, and the red region represents the facets are
sunlit, while the blue facets are referred to relatively low temperature.

\subsection{Regolith}
\begin{figure*}
\includegraphics[scale=0.9]{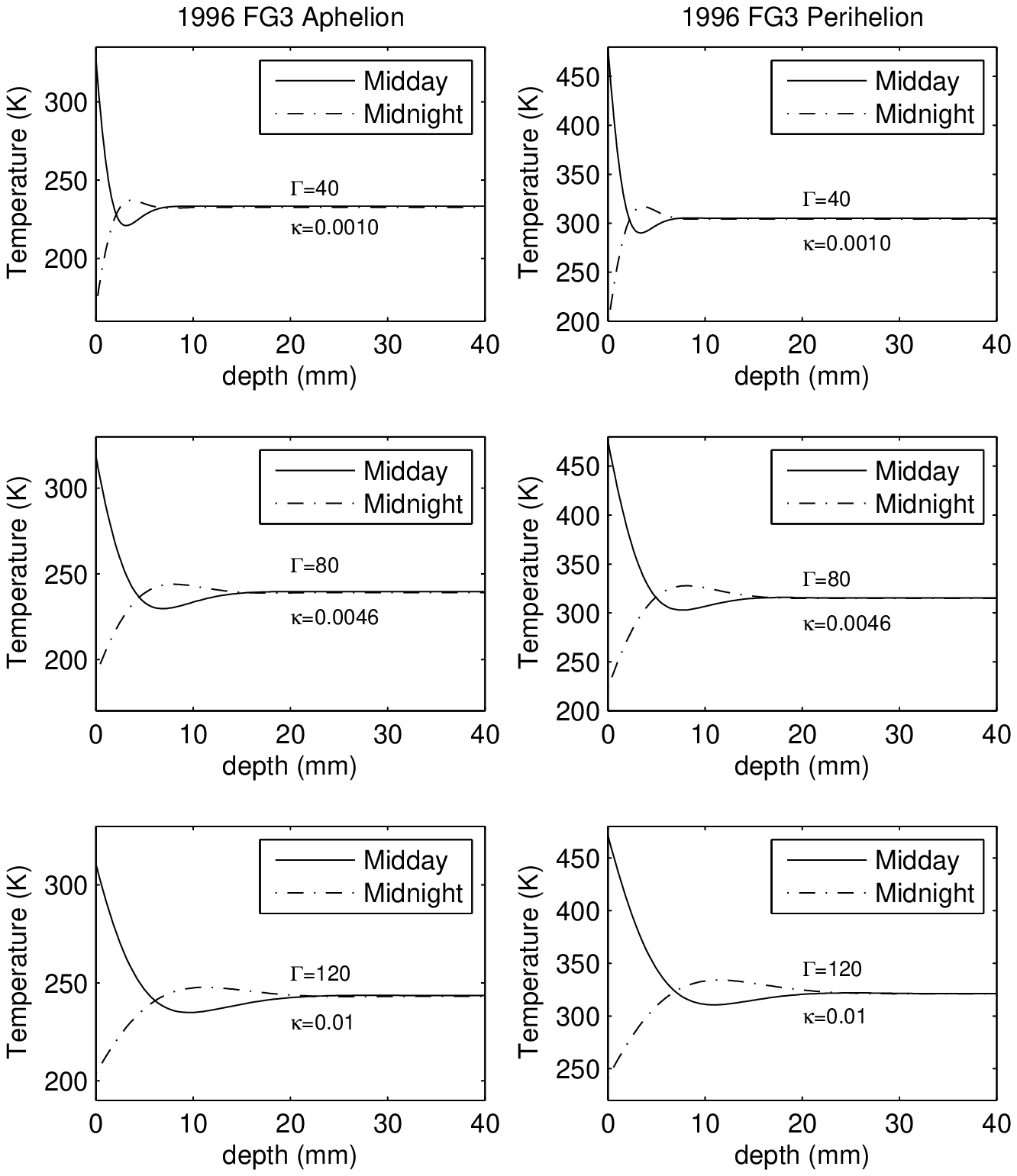}
  \centering
  \caption{Temperature distribution of sub-surface on the equator of 1996 FG3
  at local midday and midnight.
  }\label{regolith}
\end{figure*}

As mentioned previously, 1996 FG3 is chosen to be a backup target
for MarcoPolo-R sample return mission. Hence, we still show great
interest in the surface feature of the asteroid whether there
exists a regolith layer on its surface. Apparently, thermal inertia
is associated with the surface properties, where it will be helpful to
infer the presence or absence of loose material on the surface. As
known to all, fine dust has a very low thermal inertia about
$30\rm~Jm^{-2}s^{-0.5}K^{-1}$, lunar regolith owns a relatively low
value about $50\rm~Jm^{-2}s^{-0.5}K^{-1}$, a sandy regolith like Eros'
soil bears a value of $100 - 200\rm~Jm^{-2}s^{-0.5}K^{-1}$, but
coarse sand occupies a higher thermal inertia profile $\sim$
$400\rm~Jm^{-2}s^{-0.5}K^{-1}$ (e.g., Itokawa's Muses-Sea Regio). In
comparison with the above materials, bare rock has an extremely higher
thermal inertia more than $2500\rm~Jm^{-2}s^{-0.5}K^{-1}$ \citep{Delbo2007}.
In this work, the thermal inertia of 1996 FG3 is estimated to be
$80\pm40\rm~Jm^{-2}s^{-0.5}K^{-1}$. Consequently, it is quite natural
for one to suppose that the surface of 1996 FG3 may be covered by
loose materials, perhaps a mixture of dust, fragmentary rocky debris
and sand. In other word, there may exist a regolith layer on the
surface of 1996 FG3, according to the above derived thermal inertia.

Theoretically, the so-called "skin depth":
\begin{equation}
l_{s}=\sqrt{\frac{\kappa}{\rho c\omega}}
\end{equation}
is usually used to characterize the grain size of regolith.
According to the definition of thermal inertia $\Gamma=\sqrt{\rho
c\kappa}$. The profiles of mass density $\rho$, thermal conductivity
$\kappa$, and specific heat capacity $c$ may be estimated from the
above derived thermal inertia. Since the thermal conductivity
$\kappa$ depends on particle size and temperature much more than
$\rho$ and $c$, we attempt to assume a constant value for $\rho$ and
$c$, and then estimate $\kappa$ from the above derived thermal
inertia. Next, the estimation of skin depth will be easily acquired.
\citet{Sch2009} has derived the mass density to be
$1.4^{+1.5}_{-0.6}$ g cm$^{-3}$. Since the above derived value
$\Gamma=80\pm40\rm~Jm^{-2}s^{-0.5}K^{-1}$ is similar to that of the
Moon despite a little higher, we assume the specific heat capacity
of 1996 FG3 is similar to that of the Moon, about
$1000\rm~Jkg^{-1}K^{-1}$. Then the thermal conductivity $\kappa$ is
estimated to be $0.001\sim0.01\rm~Jkg^{-1}K^{-1}$ and $l_{s}$ is
$1.3\sim3.9\rm~mm$. Such small $\kappa$ and $l_{s}$ obviously
support the possible existence of loose material or regolith over
the surface of 1996 FG3. However, the depth of the regolith layer,
covered on the asteroid's surface, attracts our great attention.

We carried out simulations to explore the regolith depth
versus temperature distribution of sub-surface for
$\Gamma=40$, $80$ and $120\rm~Jm^{-2}s^{-0.5}K^{-1}$, respectively
(see Figure \ref{regolith}). The left panels show the results are
obtained when the asteroid moves at aphelion, while the right panels
exhibit those at perihelion. In each panel, two profiles, which are
respectively, plotted by solid line (local midday) and dashed line
(local midnight), show that the sub-surface temperature changes with
the depth (the distance from the surface). In these panels, the
temperature goes down as the depth increases at local noon, while
it goes up as the depth increases at local midnight until a certain
depth, where two curves are overlapped. This phenomenon results from the
internal boundary equation \ref{inbc}. The figure shows the temperature
distribution of the very loose regolith layer, the thermal conductivity
of which may be in the range of $0.001\sim0.01\rm~Jkg^{-1}K^{-1}$.
And the minimum depth of this layer may be estimated from Figure \ref{regolith}.
Herein the regolith depth of the very surface of 1996 FG3 is
reckoned to be $5\sim20\rm~mm$. This minimum value of regolith depth
may be considered as a reference for the design of a spacecraft if a
lander is equipped. On the other hand, the existence of regolith over
the surface of the primary of 1996 FG3 actually does good to a sample
return mission.

\section{Discussion and Conclusion}
In this work, we have derived a new 3D convex shape model of the
primary of 1996 FG3 from the published lightcurves, where the best-fit
orientation of its spin axis is determined to be $\lambda =237.7^\circ$ and
$\beta=-83.8^{\circ}$, with a rotation period of $\sim$ 3.5935~h.
On the basis of the numerical codes independently developed according to
thermophysical model, we apply the shape model and the required input
physical parameters to fit three sets of mid-infrared measurements for 1996 FG3.
Herein we summarize the major physical properties obtained for the asteroid as follows:
the geometric albedo and effective diameter are, respectively,
$p_{v}=0.045\pm0.002$ and $D_{\rm eff}=1.69^{+0.05}_{-0.02}\rm~km$;
the diameters of the primary and secondary are calculated to be
$D_{1}=1.63^{+0.04}_{-0.03}\rm~km$ and $D_{2}=0.45^{+0.04}_{-0.03}\rm~km$,
respectively. Moreover, the thermal inertia $\Gamma$ is also determined
to be a low value of $80\pm40\rm~Jm^{-2}s^{-0.5}K^{-1}$, whereas the roughness
fraction $f_{\rm R}$ is estimated to be $0.8^{+0.2}_{-0.4}$.

\begin{figure}
\includegraphics[scale=0.61]{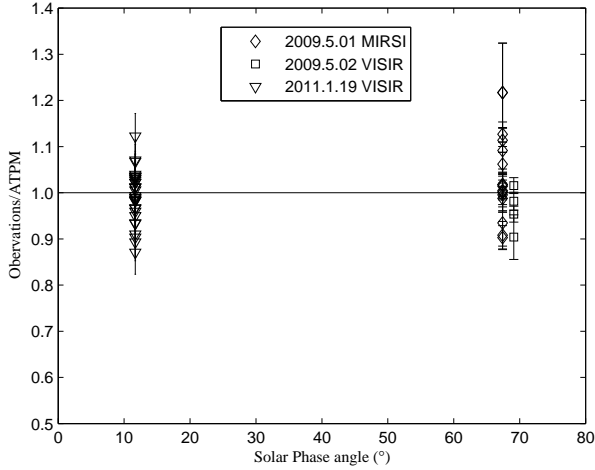}
  \centering
  \caption{The observation/ATPM ratios as a function of phase angle
  for $f_{\rm R}=0.8$, $\Gamma=80\rm~Jm^{-2}s^{-0.5}K^{-1}$,
  and $D_{\rm eff}=1.69\rm~km$.
  }\label{rphase}
\end{figure}

\begin{figure}
\includegraphics[scale=0.61]{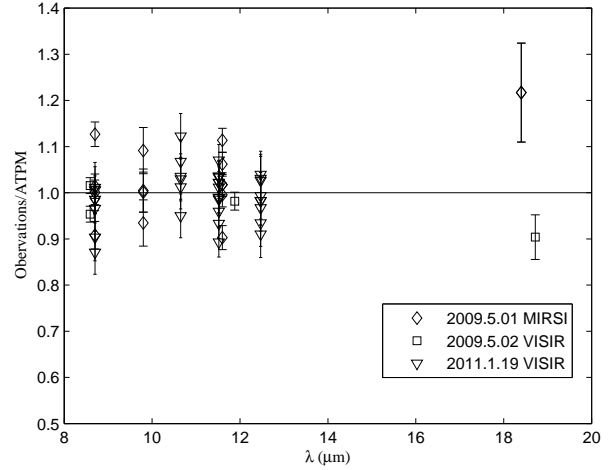}
  \centering
  \caption{The observation/ATPM ratios as a function of wavelength
  for $f_{\rm R}=0.8$, $\Gamma=80\rm~Jm^{-2}s^{-0.5}K^{-1}$,
  and $D_{\rm eff}=1.69\rm~km$.
  }\label{rspectra}
\end{figure}

From the simulations, we find that low thermal inertia
($<100\rm~Jm^{-2}s^{-0.5}K^{-1}$) would make a perfect fitting
to the VISIR observations on 19 January, 2011 (with respect to
a low phase angle), whereas the observations obtained
at higher phase angles (Table 3 and 4) are very sensitive to $f_{\rm R}$,
thereby leading to a best-fit solution with the case of large roughness fraction.
Given that we simultaneously perform the computation using the combination data
(Tables 3,4 and 5), a broad range of $L^{2}$ profiles are obtained in the fitting.
Furthermore, to acquire a significant uncertainty for the outcomes, we finally
choose a $50\%$ range of the minimum $L^{2}$ to determine the best-fit
solution for thermal inertia of 1996 FG3. On the other hand, via the
simultaneous fitting with these observations at different solar phase
angles, the degeneracy of solutions between thermal inertia and
roughness is removed, making it capable to determine the estimation
for thermal inertia and roughness separately. The ratio of "observation/model"
(see Figures \ref{rphase} and  \ref{rspectra}) is a good indicator that examines how the
results from the model match the observations at various phase angles and
wavelengths \citep{Muller2005,Muller2011,Muller2012}. Hence, this enables us to
conclude that the fitting process is correct and the derived results are
reliable.

\begin{figure*}
\includegraphics[scale=0.6]{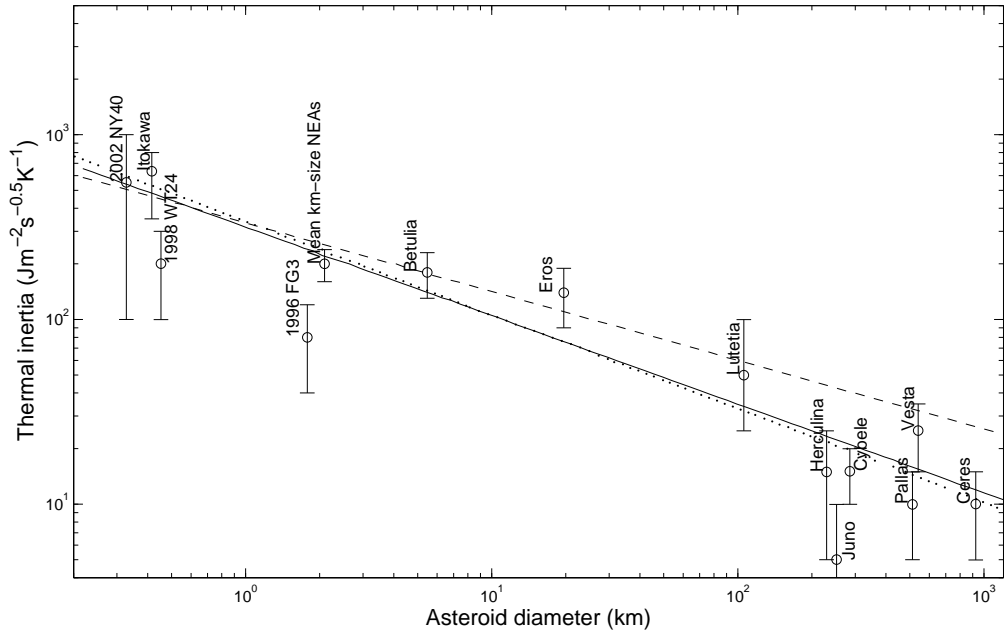}
  \centering
  \caption{Thermal inertia versus the size of asteroids \citep{Delbo2007}.
  The value of 1996 FG3 derived in this work is also labeled.
  }\label{inertia-size}
\end{figure*}

According to the ATPM fitting results shown in Table \ref{outcome},
the evaluated large roughness fraction $f_{\rm R}=0.8^{+0.2}_{-0.4}$
for 1996 FG3 may suggest a rough surface on the asteroid. However,
the asteroid's mean surface thermal inertia is estimated to be a
relative low value. Then the question arises -- why this asteroid
bears such a low thermal inertia with a very rough surface. As the
adopted roughness model in the ATPM is assumed to be an irregularity
degree of a surface at scales smaller than the global shape model
resolution (like a facet area) but larger than the thermal skin
depth.
Therefore, 1996 FG3 would be likely to have a rough surface,
which may or may not include craters of any size, and a porous or
dusty material.
The recent work of \citet{Perna2013} suggested the surface of
1996 FG3 may be a compact one, with the existence of regions
showing different roughness, similar to that of Itokawa. Hence, there would
be possible for an asteroid to possess a large roughness fraction
and low thermal inertia.

However, the NEA binaries may have a high averaged thermal inertia
of $\Gamma\approx480\pm70\rm~Jm^{-2}s^{-0.5}K^{-1}$
\citep{Delbo2011}, indicating that high thermal inertia could be
supportive to the binary formation scenario due to YORP mechanism.
In this work, we have derived thermal inertia
$\Gamma=80\pm40\rm~Jm^{-2}s^{-0.5}K^{-1}$ for 1996 FG3, and it
appears below their estimation. Nevertheless, it is still likely
for a rubble-pile asteroid to have a low thermal inertia
, thereby remaining a rough surface during a long dynamical evolution
due to space-weathering and regolith migration. In this sense, the binary
1996 FG3 may also be produced via the YORP rotation acceleration
effect at a very earlier time, but retain a roughness surface
from then, resulting in the distribution of a great many meter-size
(or even smaller) craters on the surface. Thus, these tiny craters might be
covered or surrounded by loose materials, making it appear a low thermal inertia.

\citet{Delbo2007} showed that the average thermal inertia for NEAs
may be $\Gamma=200\pm40\rm~Jm^{-2}s^{-0.5}K^{-1}$. Figure
\ref{inertia-size} exhibits the variation of mean
thermal inertia with the size of asteroids from the observations.
From this figure,  we observe that the value of thermal inertia for 1996 FG3
given in this work (labeled in red) deviates a little from the prediction profile.
In addition, our outcome for the binary system is a bit lower than
that of \citet{Wolters2011}. This results from that we perform the combination fitting
with additional thermal-IR observations (Tables 3 and 4), which may provide new insight
for the thermal study of binary asteroids.

In conclusion, the surface of 1996 FG3's primary may be a very rough surface,
on which loose materials such as fine dust, fragmentary rocky debris, sands or
most likely a stuff of their mixture are covered, composing a kind of regolith.
The depth of the possible regolith layer is evaluated to be approximately
$5\sim20\rm~mm$. Such implication may provide substantial information for
engineering of the sample return mission, for example, a selection of landing
area. However, we should place special emphasis on that this estimation is
simply a roughly minimum value of the regolith layer over the very surface
of the asteroid rather than a sort of megaregolith below the layer. Since the
thermal conductivity $\kappa$ of the megaregolith has a complicated relationship
with the depth below the surface \citep{Haack1990}, thereby we cannot simulate
the temperature distribution accurately from a one-dimension thermophysical model.
On the other hand, the formation mechanism for 1996 FG3 is still a mystery,
which the YORP acceleration mechanism may play a role in producing its shape and orbital
configuration \citep{Walsh2008}. In short, the investigations by future
space missions will throw new light on the formation scenario for this asteroid.

\section*{Acknowledgments}
The authors thank the anonymous referee and S.F. Green for their
constructive comments that significantly improve the original
contents of this manuscript. This work is financially supported by
the National Natural Science Foundation of China (Grants No.
11273068, 11203087, 10973044), the Natural Science Foundation of
Jiangsu Province (Grant No. BK2009341), the Foundation of Minor
Planets of the Purple Mountain Observatory, and the innovative and
interdisciplinary program by CAS (Grant No. KJZD-EW-Z001).

\label{lastpage}

\end{document}